\documentclass{aastex}
\usepackage{emulateapj5}

\newcommand\etal{{\it et al.\/}}
\newcommand\hkpc{\mbox{$h^{-1}\;\rm kpc$}}
\newcommand\kpc{\mbox{$\rm kpc$}}
\newcommand\bj{b$_{\rm J}$}

\begin{document}
\slugcomment{\$Revision: 1.14 $ $ accepted to {\it Astronomical Journal}}
 
\title{Shape Alignments of Satellite Galaxies}

\author{ G.~M.~Bernstein\altaffilmark{1}, P.~Norberg\altaffilmark{2}}  
\altaffiltext{1}{Department of Astronomy, University of Michigan, Ann
Arbor, MI 48109}
\altaffiltext{2}{Department of Physics, University of Durham, South 
Road, Durham DH1 3LE, United Kingdom}

\begin{abstract}
We test a sample of satellites of isolated primary galaxies, extracted 
from the 2dF Galaxy Redshift Survey (2dFGRS), for any
tendency to be aligned along (or against) the primary-to-satellite
radius vector.  If tidal effects induce such an alignment, it would
contaminate recent measurements of galaxy halo masses which use the
coherent alignment induced on background galaxies by gravitational
lensing.  The mean tangential ellipticity of 1819 satellites within
$500$~\kpc\ projected radius is
$\langle e_+ \rangle = +0.004\pm0.008,$ so no tidal alignment is
detected.  This implies at 95\% confidence that satellite
alignment is less than a 20\% contamination of the alignment signal
attributed to galaxy-galaxy lensing by Smith {\it et al.} (2001) 
and McKay {\it et al.} (2001).
\end{abstract}

\section{Introduction}
A number of the satellite galaxies of the Milky Way are distorted,
in some cases dramatically ({\it e.g.} \citet{DP01}), by the tidal
interactions with the Galaxy.  The obvious manifestations of tidal
distortion are streams of material that are liberated by tidal forces
to lead or trail the satellite in its orbit.
One might also expect the tide to stretch the satellites' gravitational
potential wells along the radial direction, and hence induce a tendency
for the bound parts of 
satellite galaxies to be extended toward their primaries. Since our
vantage point precludes a measurement of the satellites' extent along
the Galactic radius vector, it is not possible to test this
hypothesis for the Milky Way satellites.  Given a large sample of
satellites around other galaxies, however, we can test whether the
(projected) satellite major axes have a tendency to be
circumferential or radial rather than randomly oriented.  

While a detection of satellite shape alignments would
serve as a crude gauge of the effect of tidal forces on satellite
morphologies, it would be a significant nuisance for measurements of
galaxy masses using weak gravitational lensing.  In such studies, the
halo mass distribution of the primary (foreground) galaxy is inferred
by measuring the slight tendency toward circumferential alignment that
the gravitational lensing distortion of the primary galaxy induces
upon the images of distant background galaxies.  This ``galaxy-galaxy
lensing'' signal was first proposed and sought by \citet{Ty84} and
tentatively detected by \citet{Br96}.  More recently it has been
detected at high significance by \citet{F00}, \citet{Sm01}[Sm01], and
\citet{McK01}[McK01].  The latter two studies use foreground galaxies with
redshifts measured by the Las Campanas \citep{Sh96} and Sloan
\citep{Y00} redshift surveys, respectively, and are able to determine
the halo masses as a function of luminosity and morphological type.  Such
information is extremely valuable---but subject to error if the
measured mean circumferential ellipticities are due to intrinsic
alignments of the 
primaries' satellites rather than by the lensing effect.  In
galaxy-galaxy lensing measurements to date, there has been no attempt
to cull the satellites from the sample of lensed background galaxies.

In this work we test for shape alignment among a sample of satellites
of isolated galaxies found in the 2dF Galaxy Redshift Survey (2dFGRS; 
\citet{Co01}), in an effort to see if such alignments are a significant 
difficulty for galaxy-galaxy lensing studies.
The following sections describe the satellite sample; the
determination of their mean elongations; and the implications for weak
lensing studies and for tidal effects. Throughout the paper, we 
adopt a flat $\Omega_{0}=0.3$, $\Lambda_{0}=0.7$ cosmology to convert 
redshift into comoving distance and we assume the value of the Hubble 
constant to be $H_{0}=70\,{\rm km}\,{\rm s}^{-1}\,{\rm Mpc}^{-1}$.

\section{Satellite Sample}

The sample of satellite galaxies around isolated primaries is obtained 
from the 2dFGRS dataset as of December 2001. The details of how we 
identify the satellites and primaries within the survey and the tests 
on the robustness on the method used are given in \citet{No10}. However 
in order to be complete, we briefly repeat the steps of this satellite 
selection scheme below, which follows the one developed 
by \citet{Za93}. The criteria for isolation of the primary galaxy are
more relaxed in this work than for studies of satellite galaxy dynamics.
Using the full 2dFGRS redshift catalogue 
containing over 200,000 galaxies, we first select regions of sufficiently 
high redshift completeness. We then loop over all galaxies which are at 
least 2.2 magnitudes brighter than the magnitude limit of the 
2dFGRS, and localize the galaxies which are isolated. A bright 
galaxy is isolated and considered as a primary if
all its neighbouring galaxies with 
$\Delta V\,=\,|V^{\rm prim}-V^{\rm gal}|\,\le\,1000$ km/s and within 
a projected distance $\le 500$~\kpc\ are such that 
$b_{\rm J}^{\rm gal} \ge b_{\rm J}^{\rm prim} + 2.2$

Finally all galaxies around a primary which are within a projected 
distance $\le 500$~\kpc\ and with relative velocity 
$\Delta V\,=\,|V^{\rm prim}-V^{\rm gal}| \le 500$ km/s are 
considered as detected satellites around bright isolated galaxies.
Only galaxies with a good redshift measurement will be taken 
up in our catalogue of primaries and satellite galaxies. We also make 
sure, by using the full photometric 2dFGRS catalogue, that no bright 
galaxy without redshift measurement lies within the projected radius of 
an isolated primary. In other words, we eliminate any source of 
contamination due to spectroscopic determination failure.

We note here that the minimum required magnitude difference between 
primaries and satellites is about the same as the typical 
foreground/background magnitude difference in the Sm01 and McK01 lensing 
studies. This selection makes this satellite sample 
particularly appropriate for determination of the lensing contamination. 
However, we have to point out that for most primaries we have only 
detected one or two satellite galaxies, this mainly being due to the 
limiting depth of the 2dFGRS, of typically \bj=19.45. As a matter of 
fact, only 12 primaries in our catalogue of 1227 primaries have 6 or 
more galaxy satellites around them, and we have also detected 5886 
primaries without any spectroscopically confirmed satellite galaxy.
%[{\bf Peder:  need to update numbers in this last sentence.}]

\section{Ellipticity Determinations}
\subsection{Shape Measurements}
\begin{figure}
\centering
\epsscale{0.6}
\plotone{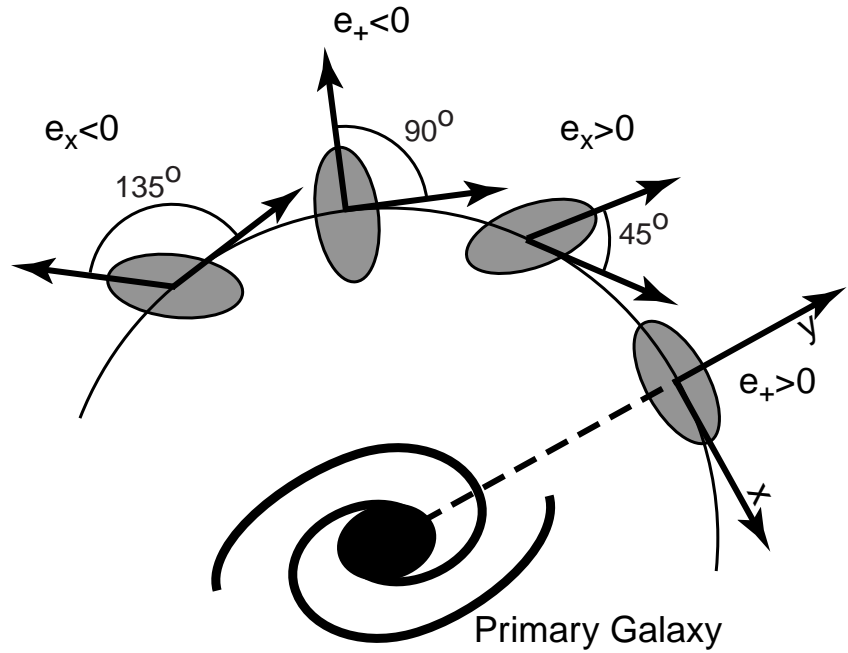}
\caption{
The orientations of satellite galaxies that would give rise to
positive and negative $e_+$ and $e_\times$ are illustrated.  The mean
value of $e_\times$ should be zero by inversion symmetry so it serves
as a systematic check.  Tidal elongations could produce a non-zero
$\langle e_+ \rangle$ among satellites, while lensing produces 
$\langle e_+ \rangle > 0$ among background galaxies.
}
\label{e1e2}
\end{figure}

Shapes for all satellites are taken from the APM scans of photographic
survey plates \citet{Ma90}. We rotate the
APM-determined shapes into a
coordinate system with the $y$ axis along the radius vector from
primary to secondary. 
There result two ellipticity components which are defined by
\begin{eqnarray}
e_+ & \equiv & { I_{xx} - I_{yy} \over I_{xx} + I_{yy}} \\
e_\times & \equiv & { 2I_{xy} \over I_{xx} + I_{yy}}
\end{eqnarray}
where $I_{ij}$ are the quadratic central moments of the galaxy
intensity integrated within a bounding isophote.  Figure~\ref{e1e2} illustrates
the meaning of $e_+$ and $e_\times$.  We expect a tidal effect to be
manifested as non-zero $\langle e_+\rangle$. 
The mean $e_\times$ must be zero if the
Universe has inversion symmetry, hence we can use $e_\times$ as a test
for systematic errors.

The unweighted isophotal moments produced by the APM software are 
noisier and more difficult to correct for point-spread-function (PSF)
distortions than more recently devised ellipticity estimators.  The
satellites are, however,  all at least 1 mag above the $b_J=20.5$ limit of the
APM catalog, and should hence have fairly high signal-to-noise and be
well resolved in the APM images.
In fact we make no correction at all for PSF effects.  Crudely,
the measured ellipticity ${\bf e}$ will be related to the true image
ellipticity ${\bf e}^i$ and the PSF ellipticity ${\bf e}^\star$ by
\begin{equation}
{\bf e} = (1-R) {\bf e}^i + R {\bf e}^\star,
\end{equation}
where $R$ is the ratio of the intensity-weighted moment $\langle r^2
\rangle$ of the PSF to that of the measured 
image.
In our application, the second term has no effect, since $\langle {\bf
e}^\star \rangle=0$ if the PSF orientation has no correlation with the
primary-satellite vector.  We can estimate the effect of the
circularization factor $(1-R)$ in the first term by noting that the
{\em measured} satellite shapes have $\langle e^2_+ \rangle
= \langle e^2_\times \rangle = (0.31\pm0.01)^2$.  This measured value 
will be a factor $(1-R)^2$ smaller than the {\em intrinsic} 
$\langle e^2_+ \rangle$ of the 
population.
\citet{BJ02}[BJ02] find the RMS intrinsic ellipticity
of high-surface-brightness galaxies in the nearby 
Universe to be 0.30.  If the intrinsic shapes of satellite are similar to
those of high-surface-brightness galaxies in an
apparent-magnitude-limited sample,
then the circularization factor $(1-R)$ is
unity to within 10\% or so. 
BJ02 find an RMS ellipticity of 0.48 for the lowest-surface-brightness 
subset of nearby galaxies.  If the satellites have this intrinsic 
dispersion, then $\langle 1-R \rangle\approx0.6$
But the higher shape dispersion of low-surface-brightness galaxies
is due to a preponderance of
disk galaxies, which should be largely absent from the satellite
sample. This prejudice is reinforced by the observation that the
distribution function for $|e|$ values for the satellites is very
similar to BJ02's distribution function for high-surface-brightness 
galaxies (their Figure~4).
It therefore seems more likely that there is negligible PSF
circularization in the APM ellipticities, so we will not make any
correction for the $1-R$ term.

As a further check on the possibility of PSF circularization, we split
the satellite galaxies into a bright half ($b_J<18.8$) and faint half
($b_J>18.8$).  The brighter half is presumably nearer and better
resolved than the fainter half, so a substantial circularization
effect would be manifested as a lower RMS ellipticity for the fainter
half.  The RMS ellipticities $\sqrt{\langle e_+^2\rangle}$ of the
brighter and fainter subsamples are $0.355\pm0.008$ and
$0.298\pm0.007$, respectively.  The circularization of the fainter
half is hence a $\approx15\%$ effect; the fainter galaxies could also
just be intrinsically rounder.

\subsection{Mean Ellipticities}
An unweighted average of the 1819 satellite ellipticities
yields
\begin{equation}
\label{emean}
\begin{array}{rcl}
\langle e_+ \rangle & = & +0.004\pm0.008 \\
\langle e_\times \rangle & = & +0.005\pm0.008.
\end{array}
\end{equation}

In the weak lensing measurements, one ideally weights low-$e$ galaxies
more heavily as they offer better sensitivity to lensing shear.  Such
a weighted average is not justified in considering the effect of tidal
forces on intrinsic satellite shapes.  It is, however, relevant for
assessing the contamination of weak lensing data by satellite
alignments since the lensing data are weighted.  Following BJ02, we
apply a weight $w(e) = (e^2+0.01)^{-1/2}$ to each satellite.  In these
scheme we obtain
\begin{equation}
\label{ewt}
\begin{array}{rcl}
\langle e_+ \rangle & = & +0.002\pm0.007 \\
\langle e_\times \rangle & = & +0.007\pm0.007.
\end{array}
\end{equation}

There is clearly no significant detection of net satellite alignment,
in either component, and the $e_\times$ values show that our uncertainties
are sensible.

\begin{figure}
\centering
\plotone{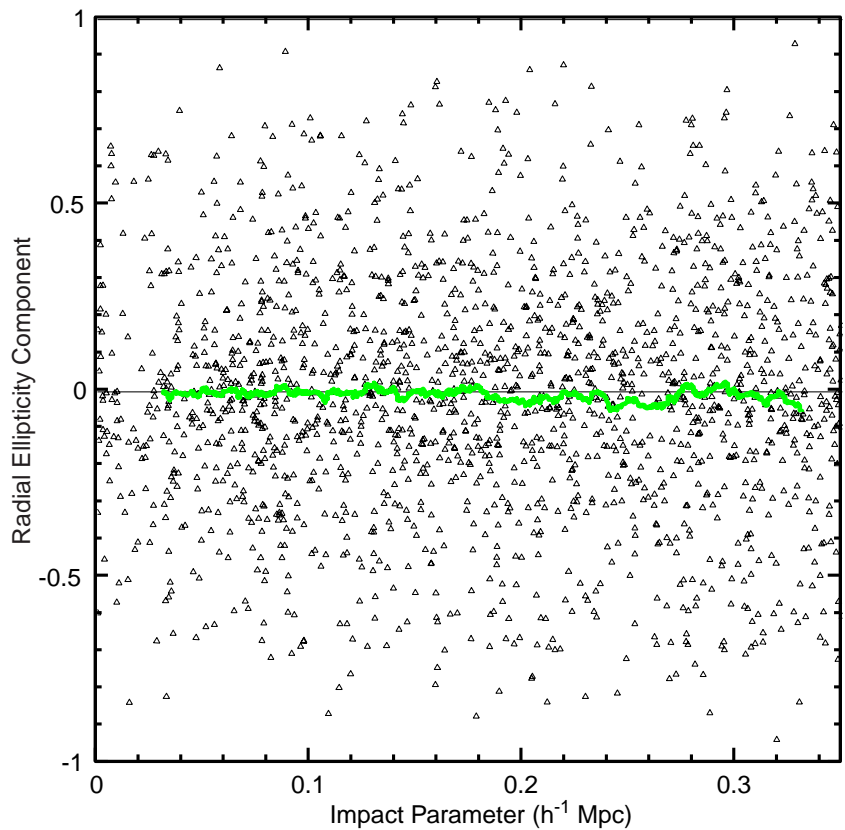}
\caption{
The satellites' $e_+$ values are plotted against their projected
distance from the primary.  The heavy (green) line plots a running 
unweighted mean of 10\% of the satellite sample.  No significant 
alignment is detected at any scale. 
}
\label{runave}
\end{figure}

As illustrated in Figure~\ref{runave}, no alignment is detected even when we
restrict the sample to the smaller projected radii where tidal effects
should be stronger.
We have also tested for alignment in the half of the galaxies that
have primaries brighter than 
$M_{\rm b_{\rm J}}=\,-21.5$.
The unweighted mean $\langle e_+ \rangle = +0.025\pm0.012$, 
a $2\sigma$ signal, but the weighted mean is only $1.4\sigma$ 
non-zero and there is no detection of real significance.  Further
restricting the sample to small projected radii does not increase the
measured mean.

\section{Implications}
\subsection{Tidal Distortions of Satellites}

Equations~(\ref{emean}) indicate that, for satellites within 
$500$~\kpc\ projected radius, the mean radial or 
circumferential elongation is less than 2\% in projection at 
95\% confidence.  We wish to place some bound on the three-dimensional
satellite shapes from the limit on the projected shape.  
We now define a parameter $\epsilon$ that characterizes the mean departure
of satellite shapes from spherical symmetry. We place a
Cartesian coordinate system on each satellite such that the $y$ axis
is along the primary-satellite radius vector, and the $z$ axis is
perpendicular to both the line of sight and the $y$ axis.  The luminosity
distribution of each satellite has second central moments $I_{xx}$,
$I_{yy}$, and $I_{zz}$ in this system.  Because the ensemble of
primary-satellite systems must be isotropic with respect to
the line of sight, the mean transverse moments must satisfy 
$\langle I_{xx} \rangle = \langle I_{zz} \rangle$.  We can then
quantify the mean tidal distortion by defining $\sigma^2$ and
$\epsilon$ such that
\begin{eqnarray}
\langle I_{xx} \rangle = \langle I_{zz} \rangle & = &
\sigma^2(1-\epsilon/2) \\
\langle I_{yy} \rangle & = &
\sigma^2(1+\epsilon) 
\end{eqnarray}
A positive value of $\epsilon$ would indicate the presence of a radial
elongation from tidal effects.  It is also possible that a nonzero
$\epsilon$ could arise if satellites tend to be found along and align
with the dark matter filaments that are believed to underly the
large-scale distribution of galaxies.  We are not aware of any
theoretical estimates of the size (or sign!) expected of such an
effect.

If there is no absorption in the galaxy, then the mean {\em projected}
second moments can be simply related to the intrinsic elongation
$\epsilon$ via
\begin{equation}
\langle e_+ \rangle = \epsilon/2.
\end{equation}
We may thus bound the mean radial asymmetry of satellites to 
be $\langle \epsilon \rangle < 4\%$ at 95\% CL.

\subsection{Weak Lensing Measurements}
Equations~(\ref{ewt}) imply that the population of satellites will
produce a false lensing distortion of $|\delta|<0.016$ at 
95\% CL.
F00 measure the angular correlation between foreground and ``background''
samples and infer that the pollution of the ``background'' sample by
neighbors of the foreground galaxy---{\it i.e.} satellites---is about
15\% within impact parameters of 60~\hkpc.  The contamination in Sm01
is significantly lower, 10\% at most.  Hence the deleterious effect of
satellite 
alignment will be significantly diluted at these radii, and will be
even weaker at larger radii.  For the Sm01 data, we may conclude that
the false distortion signal produced by aligned satellites satisfies
\begin{equation}
|\delta_{\rm false}|<0.0016
\end{equation}
at $95\%$ confidence within 60~\hkpc\ impact parameter.  Since Sm01
detect a mean distortion of $\delta\approx0.008$ at this impact
parameter around an $L_\star$ galaxy at $z=0.1$, the error due to
satellite alignments is at most a 20\% effect.  We cannot yet
conclude that the effect is totally negligible, but this is likely the
case, especially at larger impact parameters, where the satellite
fraction is smaller, and the tidal effect drops more rapidly than the
lensing shear.

Larger and deeper satellite samples containing more primaries with 
many satellite galaxies each may reveal a significant alignment, and 
future galaxy-galaxy lensing measurements
will be aiming for higher precision.  If the satellite alignments do
ultimately pose a barrier to high precision, then their deleterious
effect could be eliminated by using photometric redshift
estimates to exclude the bulk of the satellite population from the
sample of lensed background galaxies. Satellite alignments do not,
however, seem to be a problem for current galaxy-galaxy lensing data.  

\acknowledgements
We are grateful to the 2dFGRS team for allowing us to use their data prior
to publication and to Carlos Frenk for useful discussions. 
The 2dFGRS is being carried out using the 2 degree field facility on the 
3.9m Anglo\-Australian Telescope (AAT). We thank all those involved in 
the smooth running and continued success of the 2dF and the AAT.
This work was supported in part by grant AST-9624592 from the 
National Science Foundation and in part by a PPARC rolling grant 
at Durham. PN is supported by the Swiss National Science 
Foundation and an ORS award.

\newpage
%\figcaption[sate1e2.eps]

%\figcaption[runave.eps]{


\begin{thebibliography}{}

\bibitem[Bernstein \& Jarvis (2002)]{BJ02}
Bernstein, G. M., \&  Jarvis, M.  2002, \aj, 123, 583

\bibitem[Brainerd, Blandford, \& Smail (1996)]{Br96} 
Brainerd, T.G., Blandford, R.D., \& Smail, I. 1996, \apj, 466, 623 

\bibitem[Colless \etal (2001)]{Co01} 
Colless, M.M., \etal (the 2dFGRS Team), 2001, MNRAS 328, 1039 

\bibitem[Dohm-Palmer \etal (2001)]{DP01}
Dohm-Palmer, R., \etal\ 2001, \apj, 555, 37

\bibitem[Fischer \etal (2000)]{F00}
Fischer, P. \etal\ 2000 , \aj, 120, 1198

\bibitem[Maddox \etal (1990)]{Ma90}
Maddox, S.J., Sutherland, W.J., Efstathiou, G., \& Loveday, J, 
1990, MNRAS 243, 692

\bibitem[McKay \etal (2001)]{McK01}
McKay, T. \etal\ 2001 , \apj, (submitted) 
(astro-ph/0108013)

\bibitem[Norberg \etal (in prep.)]{No10}
Norberg, P.,  \etal, in preparation

\bibitem[Shectman \etal (1996)]{Sh96}
Shectman, S. A., Landy, S. D., Oemler, A., Tucker, D. L., Lin, H., Kirshner, R.P., \& Schechter, P. L. 1996, \apj, 470, 172

\bibitem[Smith \etal (2001)]{Sm01}
Smith, D. R., Bernstein, G. M., Fischer, P., \& Jarvis, M. 2001, \apj,
551, 643

\bibitem[Tyson \etal (1984)]{Ty84}
Tyson, J.A, Valdes, F., Jarvis, J.F., \& Mills, A.P. 1984, \apjl, 281, L59

\bibitem[York \etal (2000)]{Y00}
York, D. \etal\ 2000, \aj, 120, 1579

\bibitem[Zaritsky \etal (1993)]{Za93}
Zaritsky, D., Smith, R., Frenk, C., \& White, S.D.M., 1993, \apj, 405, 464

\end{thebibliography}
\end{document}